\documentclass[10pt, twocolumn]{IEEEtran}

\usepackage{cite}   % Written by Donald Arseneau

\usepackage{graphicx}  % Written by David Carlisle and Sebastian Rahtz
\usepackage{amsmath, amsopn, psfrag, amsthm}   % From the American Mathematical Society
\usepackage{amssymb}
\usepackage{color}
\usepackage{algorithm}
\usepackage{algorithmic}
\usepackage[utf8]{inputenc}
\usepackage[english]{babel}

\newtheorem{lemma}{Lemma}

\usepackage{psfrag}
\usepackage{epsfig}

\usepackage{subcaption}
\usepackage{kantlipsum} %<- For dummy text
\usepackage{mwe} %<- For dummy images
\DeclareCaptionLabelSeparator{periodspace}{.\quad}
\usepackage{amssymb,amsmath,mathtools}

\usepackage{cite}   % Written by Donald Arseneau
\usepackage{graphicx}  % Written by David Carlisle and Sebastian Rahtz
\usepackage{amsmath, amsopn}   % From the American 
\usepackage{amssymb}
\usepackage{color}
\usepackage{psfrag}
\usepackage{epsfig}
\usepackage{stfloats}
\usepackage{lipsum}
\usepackage{multicol}

\def\bfphi{{\boldsymbol{\phi}}}
\def\bftheta{{\boldsymbol{\theta}}}

\graphicspath{{./img/}}
\DeclareGraphicsExtensions{.pdf,.jpeg,.png, .eps}
\DeclareMathOperator*{\argmax}{arg\, max}

\usepackage{amssymb,amsmath,mathtools}

\begin{document}
\setlength{\columnsep}{.1in}

\title{Evaluation of Low Complexity Massive MIMO Techniques Under Realistic Channel Conditions}
\author{
Manijeh Bashar,~\IEEEmembership{Student Member,~IEEE}, Alister G. Burr,~\IEEEmembership{Senior Member,~IEEE},  Katsuyuki Haneda,~\IEEEmembership{Member,~IEEE}, Kanapathippillai Cumanan,~\IEEEmembership{Member,~IEEE}, Mehdi M. Molu,~\IEEEmembership{Member,~IEEE}, Mohsen Khalily,~\IEEEmembership{Senior Member,~IEEE} and Pei Xiao,~\IEEEmembership{Senior Member,~IEEE
} 

\thanks{M. Bashar, A. G. Burr, and K. Cumanan  are with the Department of Electronic Engineering, University of York, Heslington, York, U.K. Email: \{mb1465, alister.burr, kanapathippillai.cumanan\}@york.ac.uk. M. Bashar is also with
	 home of the 5G Innovation Centre, Institute for Communication Systems, University of Surrey, U.K. e-mail: m.bashar@surrey.ac.uk.} 
\thanks{K. Haneda is with Aalto
University School of Electrical Engineering, Espoo, Finland. Email: {katsuyuki.haneda@aalto.fi}}
\thanks{M. M. Molu is with Samsung Cambridge Solution Centre
(SCSC), UK, Email: mehdi.molu@gmail.com.}
\thanks{M. Khalily and P. Xiao are with the 5G Innovation
Center, Institute for Communication Systems, University of Surrey, Guildfordm U.K. Email: \{m.khalily, p.xiao\}@surrey.ac.uk.}
\thanks{This work was supported by H2020- MSCA-RISE-2015 under grant number 690750. Moreover, the work on which this paper is based was carried out in collaboration with COST Action CA15104 (IRACON).}
\thanks{K. Haneda would like to acknowledge the financial support from the Academy of Finland research project ``Massive MIMO: Advanced Antennas, Systems and Signal Processing at mm-Waves (M3MIMO)'', decision $\#288670$.}
		\thanks{The work of P. Xiao was supported in part by the European Commission under the 5GPPP project 5GXcast
(H2020-ICT-2016-2 call, grant number 761498) as well as by the U.K. Engineering and Physical Sciences  Research  Council  under Grant EP/ R001588/1.
} 
}

\markboth{IEEE TRANSACTIONS ON VEHICULAR TECHNOLOGY,~2019}
{Shell \MakeLowercase{\textit{et al.}}: Bare Demo of IEEEtran.cls for IEEE Communications Society Journals}
\maketitle

%%%%%%%%%%%%%%%%%%%%%%%%%%%%%%%%%%%%%%%%%%%%%%%%%%%%%%%%%%%%%%%%%%%%%%%%%%%%%%%%%%%%%%%%%%%%%%%%%%%%%%%%%%%%%%%%%%%%%%%%%%%%%%%%%%%%%%%%%%%%%%%%%
% multiple-input multiple-output (MIMO)
% channel state information at the transmitter (CSIT)
% channel state information (CSI)
% minimum mean square error (MMSE)
% Geometry-based stochastic channel models (GSCMs)
% multi-user (MU)
% wide sense stationary (WSS)
% angle of departure (AoD)
% multi-path components (MPCs)
% joint spatial division and multiplexing (JSDM)
% greedy weight clique (GWC)
% Correlation-based user scheduling and beamforming (CUSBF)
% time division duplexing (TDD)
% additive white Gaussian noise (AWGN)
% direction of departure (DoD)
% direction of arrival (DoA) 
% base station (BS)
% mobile station (MS)
% visibility region (VR)
% non-line-of-sight (NLoS)
% cumulative distribution function (CDF)
% zero-forcing beamforming (ZFBF)
%%%%%%%%%%%%%%%%%%%%%%%%%%%%%%%%%%%%%%%%%%%%%%%%%%%%%%%%%%%%%%%%%%%%%%%%%%%%%%%%%%%%%%%%%%%%%%%%%%%%%%%%%%%%%%%%%%%%%%%%%%%%%%%%%%%%%%%%%%%%%%%%%
\begin{abstract}
A low complexity massive multiple-input multiple-output (MIMO) technique is studied with a geometry-based stochastic channel model, called COST 2100 model. 
We propose to exploit the discrete-time Fourier transform of the antenna correlation function to perform user scheduling. The proposed algorithm relies on 
a trade off between the number of occupied bins of the eigenvalue spectrum of the channel covariance matrix for each user and spectral overlap among the selected users. We next show that linear precoding design can be performed based only on the channel correlation matrix.
The proposed scheme exploits the angular bins of the eigenvalue spectrum of the channel covariance matrix to build up an ``approximate eigenchannels'' for the users. We investigate the reduction of average system throughput with \textit{no channel state information at the transmitter (CSIT)}. Analysis and numerical results show that
while the throughput slightly decreases due to the absence of CSIT, the complexity of the system is reduced significantly.
\\
$~~${$~~${\textbf{$~~$\textit{$~\textcolor{white}{cc}~$Index terms}}}— COST 2100 channel model, massive MIMO, MMSE estimation, spatial correlation, user scheduling.}
\end{abstract}
\section{Introduction}
Massive multiple-input multiple-output (MIMO) is a technology which involves an increased number of base station (BS) antennas and users in a multi-user (MU)-MIMO system.
To investigate the performance of massive MIMO systems, an accurate multi-user channel model is necessary. Most standardized MIMO channel models such as IEEE $802.11$, the 3GPP spatial model, and the COST 273 model rely on clustering \cite{standard}. Geometry-based stochastic channel models (GSCMs) consider the physical reality of channels to investigate the performance of MIMO systems using the concept of clusters \cite{Molish_tufvesson}. The COST 2100 model is a well known GSCM \cite{Molish_tufvesson,ourglobecom_cost}.

In massive MIMO, a very large number (hundreds or even thousands) of antennas communicate with a large number of users, where the number of users is much smaller than the number of BS antennas.
Hence, an important issue in massive MIMO systems is user scheduling \cite{CiareJointspatial,cairespatial13inftheory,Yang_spawc_massive_user} in which multiuser diversity gain is obtained with imperfect channel state information (CSI).
Recently, a range of user scheduling schemes have been proposed for large MIMO systems. Most of these, such as that described in \cite{Lee14user}, require accurate knowledge of the channel from all potential users to the BS, which in the massive MIMO case is completely infeasible to obtain. However, a simplified correlation-based user scheduling algorithm, is still an open problem.

The problem of correlation-based user scheduling and precoding in cluster-based channel models and its effect on the system performance of massive MIMO has not been well studied in the literature. 
In \cite{hanzo_mm}, the authors assume that each scattering cluster contributes
a single propagation path. However, \cite{cairespatial13inftheory} considers the cluster-based millimetre-wave (mm-wave) channel model, and investigates the effect of shared clusters on the system performance.
Note that in this work, the effect of shared and distinct clusters on the system throughput is considered. Moreover, in \cite{Spencer_zf_04}, the authors consider multi-antenna receivers and exploit block-diagonalization, which is a generalization of channel inversion when there are multiple antennas at each receiver whereas in this paper, we have considered single-antenna receivers. 
Coordinating receiver antennas through receiver processing is still beneficial for a finite number of antennas \cite{Spencer_zf_04}.
Interestingly, however, in \cite{SUSGoldsmithGlobcom,goldsmith_jsac3} the authors show that the asymptotic performance as the number of users tends to infinity is not improved by antenna cooperation.
In \cite{ouriet_mic,our_ew}, the authors present a robust user selection algorithm based on knowledge of the geometry of the service area and of location of clusters, without having full CSI at the BS. The problem of user scheduling with delayed channel is investigated in \cite{ouriet_eslami,ourvtc_eslami}.

In this paper, we investigate the problem of joint user scheduling and beamforming design when only knowledge of the statistics of the channel is available at the BS.
The second order statistics of the channel depend on the position of the users and the geometry of the system, including the relative position of clusters in the area with respect to the BS and users. The fixed positions of the users and clusters mean that a wide sense stationary (WSS) process is an appropriate model for the statistics of the channel. In the other words, if the geometry of the system is fixed, the channel covariance matrix remains constant over time. Moreover, changing the position of the users by a few meters will not affect the statistics of the channel \cite{CiareJointspatial}. 

In general, multi-path components (MPCs) from shared clusters cause correlation which reduces the rank of the channel \cite{burrijas,alister_iswcs}. We therefore work in this paper on the effect of shared bins on the system performance. 
Given the second order statistics of the channel, we perform low-complex user scheduling and precoding based only on the covariance matrix of the users. 
The behaviour of the eigenvalues of the channel covariance matrix for a large number of antennas at the BS is studied. When the number of antennas tends to infinity, based on Szego's theorem for large Toeplitz matrices \cite{CiareJointspatial,szego}, the eigenvalue spectrum of the channel covariance matrix can be obtained by the discrete-time Fourier transform of the antenna correlation function. 
In this paper, we assume that the carrier frequency is 2 GHz and hence improve the existing design method of the beamformer matrix for the mm-wave range \cite{cairespatial13inftheory,CiareJointspatial,caire_ref_mmwave} so that the method takes advantage of the nature of clustered channels at 2 GHz. The existing methods \cite{cairespatial13inftheory,CiareJointspatial,caire_ref_mmwave} are based on an assumption that the spatial multipath channels are sparse, while the assumption is not valid at 2 GHz and hence the design method is not directly applicable. Furthermore, we for the first time design a beamformer matrix for the COST 2100 channel model that is parameterised based on extensive urban MIMO measurements.

Massive MIMO is attractive in the range of 1.5-3 GHz band from the perspective that considering half-wavelength spacing, the authors in \cite{Chien_emil_bookchapter} emphasise that an array area
of 1 $\text{m}^2$ can accommodate 100 antennas at a 1.5 GHz carrier frequency and 400 antennas at 3 GHz. \cite{Chien_emil_bookchapter, emil_myth_magazin_16}. Our results and contributions are summarized as follows:
\begin{itemize}
\item[\textbf{1.}] Exploiting the eigenvalue spectrum of the channel covariance matrix, we propose to use the angular bins to build up an approximate eigenchannel, which can be used for linear precoding design. Next, a new user scheduling scheme is proposed \textit{under the assumption that no instantaneous channel information is available at the BS, other than the channel correlation}.
\item[\textbf{2.}]The complexity of different schemes is investigated. Moreover, we show that exploiting the proposed algorithm, the computational complexity of the system reduces significantly.
\item[\textbf{3.}] Numerical results show significant performance improvement compared to the joint spatial division and multiplexing (JSDM)-based scheduling scheme presented in \cite{CiareJointspatial}. Moreover, in \cite{SUSGoldsmithGlobcom}, the BS exploits knowledge of the estimated channel to design the beamformer. Hence it is very difficult to achieve the performance of the greedy weight clique (GWC) scheme \cite{SUSGoldsmithGlobcom} knowing only the correlation matrix. The numerical results confirm that there is only a small gap (5-8 bits/s/Hz in terms of achievable sum rate) between the performance of the proposed correlation-based scheme and the GWC scheme (which relies on the availability of the channel estimates at the BS).
\end{itemize}  
\section{SYSTEM MODEL}
Consider downlink transmission in a single cell with $M$ antennas at the BS and $K$ single antenna user terminals on the same time-frequency resource. Here, we assume time division duplexing (TDD) mode where the uplink and downlink channel are the same. 
\subsection{Downlink Transmission}
The transmitted signal when $K_s\, (K_s\ll M)$ users have been selected from the pool of $K$ users, is given by 
$
\textbf{x}= \sum_{k=1}^{K_s}{\sqrt{p_{k}}\textbf{w}_{k}{s} _{k}},
$
where ${s}_{k}$ denotes the data symbol of user \textit{k}, $\textbf{w}_{k}$ denotes the precoding vector of size $M$ and $p_{k}$ denotes the power assigned to user $k$. Then the received signal at user $k$ is given by
\begin{IEEEeqnarray}{rCl}
y_{k}&=&\sqrt{p_{k}}\textbf{h}_{k}\textbf{w}_{k}{s}_{k}+\sum_{j=1,j\ne k}^{K_s}{\sqrt{p_{j}}\textbf{h}_{k}\textbf{w}_{j}{s}_{j}}+n_{k},
\label{forml2}
\end{IEEEeqnarray}
where the vector $\textbf{h}_{k}$ of size $M$ denotes the downlink channel of the $k$th ($k=1,\cdots,K_s$) user and $n_{k} \in \mathbb{C}(0,1)$ is the complex additive white Gaussian noise (AWGN) element. 
\subsection{Geometry-based Stochastic Channel Model}
GSCMs are mathematically tractable
models to investigate the performance of MIMO systems
\cite{Molish_tufvesson}, where the double directional channel impulse response is a superposition of MPCs as given by \cite{Costaction,Molish_tufvesson}
\begin{equation}
h(\!t\!,\!\tau\!,\!\bfphi^{\text{BS}}\!,\!\bftheta^{\text{MS}}\!)\!=\!\!\sum_{j=1}^{N_C}\!\sum_{i=1}^{N_p}\!\!a_{i,j}\delta(\!\phi^{\text{BS}}-\phi_{i,j}^{\text{BS}})\delta(\!\theta^{\text{MS}}\!-\!\theta_{i,j}^{\text{MS}})\delta(\tau-\tau_{i,j}\!),\!
\small
\label{h1}
\end{equation}
where $N_p$ and $N_C$ are the number of MPCs and the total number of clusters, respectively, $t$ is time, $\tau$ denotes the delay, $\delta$ denotes the Dirac delta function, and $\bfphi^{\text{BS}}$ and $\bftheta^{\text{MS}}$ represent the direction of departure (DoD) at the BS and direction of arrival (DoA) at the mobile station (MS), respectively. Similar to \cite{Costaction}, we group the MPCs with similar delay and directions into clusters. The circular visibility region (VR) determines whether the cluster is active or not for a given user. The MPC's gain scales by a transition function of the VR that is given by $A_{\text{VR}}$ and is a function of the distance between the MS and the VR centre \cite{Costaction}.
We assume Rayleigh fading for the MPCs within each cluster. The complex amplitude of the $i$th MPC in the $j$th cluster in (\ref{h1}) is given by
 \begin{equation}
a_{i,j}=L_pA_{VR}\sqrt{A_C A_{\text{MPC}}},
\label{a}
\end{equation} 
where $L_p$ is the channel path loss, $A_{\text{MPC}}$ is the Rayleigh-faded power of each MPC, and $A_C$ refers to the cluster power attenuation \cite{Costaction}. For the non-line-of-sight (NLoS) case of the micro-cell scenario, the path loss is
$
L = 26 \log_{10}d_{\text{BS,MS}}+20\log_{10}(4\pi/\lambda),
$
where $d_{\text{BS,MS}}$ and $\lambda$ denote the distance from the BS to the MS and the wavelength {in meters}, respectively.
\section{Eigenvalue spectrum of the antenna correlation function}
In the COST 2100 channel model, each entry of the channel matrix can be written as
\begin{IEEEeqnarray}{rCl}
h_{km}=\sum_{i=1}^{N_l}
a_{ki}
~\delta(\phi-\phi_{ki})\delta(\theta-\theta_{ki})\delta(\tau-\tau_{ki}),
\label{cost}
\end{IEEEeqnarray}
where $N_l=N_C\times N_p$, and it denotes the total number of paths and $\bfphi_{ki}$ and $\bftheta_{ki}$ represent the DoD and DoA respectively of path $i$ to the $k$th user.
The complex amplitude of the $i$th MPC in (\ref{cost}) is given by
\begin{IEEEeqnarray}{rCl}
a_{ki}\!=\!\underbrace{L_p~A_{\text{VR}}~\sqrt{A_C}}_{\text{geometry-based attenuation}} \!\times\!\underbrace{A_{\text{MPC}}}_{\text{small-scale fading}}
\!=\!a_{ki}^{\text{ga}}\times a_{ki}^{\text{sf}}.
\label{a}
\end{IEEEeqnarray}
Note that the power of each path in (\ref{a}) is scaled with respect to the small-scale fading and the attenuation due to the geometry of the system which we call \textit{geometry-based attenuation}. Hence, assuming a linear array response at the BS side the $K\times M$ aggregate channel of all $K$ users is given by
\begin{equation}
\small
\!\!\textbf{H}\!=\!\!
\begin{bmatrix} 
\!\sum_{i=1}^{N_l}a_{1i} & \!\!\!\!\!\!\sum_{i=1}^{N_l}a_{1i}e^{j\alpha\sin\phi_{1i}}  & \!\!\!\!\!\!\ldots & \!\!\!\!\!\!\!\!\sum_{i=1}^{N_l}a_{1i}e^{j\alpha(M-1)\sin\phi_{1i}}\!\! \\
\!\sum_{i=1}^{N_l}a_{2i} & \!\!\!\!\!\!\sum_{i=1}^{N_l}a_{2i}e^{j\alpha\sin\phi_{2i}} & \!\!\!\!\!\!\ldots & \!\!\!\!\!\!\!\!\sum_{i=1}^{N_l}a_{2i}e^{j\alpha(M-1)\sin\phi_{2i}}\!\! \\
\!\!\!\vdots & \!\!\!\!\!\!\!\!\vdots & \!\!\!\!\!\!\!\!\ddots & \!\!\!\!\!\!\!\!\vdots\!\! \\
\!\sum_{i=1}^{N_l}a_{Ki} & \!\!\!\!\!\!\sum_{i=1}^{N_l}a_{Ki}e^{j\alpha\sin\phi_{Ki}} & \!\!\!\!\!\!\ldots & \!\!\!\!\!\!\!\sum_{i=1}^{N_l}a_{Ki}e^{j\alpha(M-1)\sin\phi_{Ki}}\!
\!\end{bmatrix}\!\!\!,
\label{h3}
\end{equation}
where $\alpha=-2\pi\frac{d}{\lambda}$, $d$ is the spacing between two antenna elements and $\lambda$ denotes the wavelength {(in m)}.
The $M \times M$ channel spatial covariance of the $k$th user 
channel vector is given by $\textbf{R}_{k}=\mathbb{E}\{\textbf{h}_k^H\textbf{h}_k\}$.
Assuming that the positions of users and clusters are fixed, the expectation is taken over the power of MPCs which have the Rayleigh fading distribution. Assuming a linear array response for the AoD $\phi$ and WSS over the array, each $(m,n)$-th entry of the channel covariance matrix for the $k$th user, $\textbf{R}_k$, is given by
$
[\textbf{R}_k]_{m,n}=
 \sum_{i=1}^{N_l}\!\left(a_{ki}^{\text{ga}}\right)^2e^{j\alpha(n-m)\sin\phi_{ki}},
$
where the second equality comes from the fact that $\mathbb{E}\left\{|a_{ki}|^2\right\}=(a_{ki}^{\text{ga}})^2$.
$\!\!\!\!\textcolor{white}{a}^1$\footnote{$\textcolor{white}{a}^1$Note that the measurement results in \cite{iraconmeeting_correlation} show that at the frequency of 2 GHz, to calculate the channel covariance matrix, the BS needs to average the channel samples over around 300-400 samples and 100-200 samples for the case of urban and rural environments, respectively.}
\subsection{Eigenvalue Spectrum with $M\to \infty$}
In \cite{cairespatial13inftheory}, the authors exploit Szego's theory for large Toeplitz matrices \cite{szego}, and show that for massive MIMO systems, the eigenvalue spectrum of the antenna correlation function converges to
the discrete-time Fourier transform of the antenna correlation
function. In other words, in the limit of a large number of antennas, the empirical eigenvalue cumulative distribution function (CDF) of the empirical eigenvalues from the channel correlation matrix can be approximated by the samples of the discrete-time Fourier transform of the antenna correlation
function \cite{cairespatial13inftheory}.
The eigenvalue spectrum, $S_k(f)$, is obtained by the discrete-time Fourier transform of the autocorrelation function. Hence, we consider the spectrum over the range $f \in[\dfrac{-1}{2},\dfrac{1}{2}]$. As the eigenvalue spectrum can take any positive real value, similar to \cite{cairespatial13inftheory}, we write $S_k(f); f \in [\dfrac{-1}{2},\dfrac{1}{2}] \to \mathbb{R}^+$, where $\mathbb{R}^+=\{x\in \mathbb{R}| x>0\}$ refers to the positive real values.
Each entry of the channel correlation matrix for the $k$th user is given by $r_{k(mn)} = [\textbf{R}_k]_{m,n}$, which with a change of notation, we rewrite as $r_{k(m)} = [\textbf{R}_k]_{l,l-m}$. Hence, the general expression for the discrete-time Fourier transform of the antenna correlation
function is given by the following Lemma.
\begin{lemma}
The discrete-time Fourier transform of the antenna correlation for COST 2100 channel model with large number of antennas at the BS is obtained as:
\begin{IEEEeqnarray}{rCl}
S_k(f)&=& \sum_{m=-\infty}^{\infty}r_{k(m)}e^{-j2\pi fm}
\nonumber
\\
	&=&
	\sum_{m=-\infty}^{\infty}\left(\sum_{i=1}^{N_l}\left(a_{ki}^{\text{ga}}\right)^2e^{-j2\pi\frac{d}{\lambda}(m)\sin\phi_{ki}}\right)e^{-j2\pi fm}
	\nonumber
	\\
	&&  \: \stackrel{(a)}{=}\sum_{i=1}^{N_l}(a_{ki}^{\text{ga}})^2\sum_{m=-\infty}^{\infty}\delta\left(m-\left(\frac{d}{\lambda}\sin\phi_{ki}+f\right)\right)\nonumber\\
	&&  \: =\sum_{i=1}^{N_l}(a_{ki}^{\text{ga}})^2\delta\left(f+\frac{d}{\lambda}\sin\phi_{ki}\right),
	\label{s1}
\end{IEEEeqnarray}
where the step $(a)$ comes from the property of sum of complex exponentials \cite{CiareJointspatial}.
\end{lemma}
Equation (\ref{s1}) shows that the DoD of paths can be estimated perfectly from  the eigenvalue
spectrum in the case of $M\to \infty$. In the next section, we show that the eigenvalue spectrum of $\textbf{R}_k$ can be used to build up an approximate eigenchannel matrix for precoding and user scheduling.
\subsection{Eigenvalue Spectrum with Finite $M$}
For the case of finite $M$, this paper follows the methodology in \cite{CiareJointspatial}. In \cite{CiareJointspatial}, Adhikari et al. proposed quantizing the interval $[-\frac{1}{2},\frac{1}{2}]$ into $M$ disjoint intervals of size $\frac{1}{M}$. Using analysis in \cite{CiareJointspatial}, each interval introduces an angular bin, where bin $B_b$ is centred at 
$\frac{b}{M}-\frac{1}{2}$ with $b\in \{0,1,\cdots,M-1\}$. Hence, based on \cite{CiareJointspatial}, the \textit{k}th user ``occupies'' bin $B_b$ if the following condition holds:
\begin{IEEEeqnarray}{rCl}
\small
\!\!-\frac{d}{\lambda} \sin \phi_{kp} \!\in\! B_b \!\equiv\! \!\frac{b}{M}\!-\!\frac{1}{2}\!-\!\frac{1}{2M}\!\!<\!\!\!-\!\frac{d}{\lambda} \sin \phi_{kp} \!\! \le\!\! \frac{b}{M}\!-\!\frac{1}{2}\!\!+\!\!\frac{1}{2M}\!.~~~~
\end{IEEEeqnarray}
Let us assume, similar to \cite{CiareJointspatial}, that $\pi(i)$ denotes the index of
the bin occupied by the MPC $i$. Then, based on \cite{CiareJointspatial}, $S_k(f)$ for the case of finite $M$ can be written as
\begin{IEEEeqnarray}{rCl}
S_k(f)&=&\sum_{i=1}^{N_l}(a_{ki}^{\text{ga}})^2\sum_{m=-\infty}^{\infty}\delta\left(m-\left(\frac{d}{\lambda}\sin\phi_{ki}+f\right)\right)\nonumber\\
 &=&\sum_{i=1}^{N_l}(a_{ki}^{\text{ga}})^2 \times 1 \left\{f \in B_{\pi(i)}\right\}.
\label{s2}
\end{IEEEeqnarray}
\begin{figure*}[!b]

\hrulefill

\begin{IEEEeqnarray}{rCl}
\begin{split}
\textbf{U}\!=\!
\begin{bmatrix} 
\sum_{i,-\frac{d}{\lambda} \sin \phi_{1i} \in B_1}^{}(a_{1i}^{\text{ga}})^2 
&
\sum_{i,-\frac{d}{\lambda} \sin \phi_{1i} \in B_2}^{}(a_{1i}^{\text{ga}})^2
&\ldots & 
\sum_{i,-\frac{d}{\lambda} \sin \phi_{1i} \in B_M}^{}(a_{2i}^{\text{ga}})^2
 \\
\sum_{i,-\frac{d}{\lambda} \sin \phi_{2i} \in B_1}^{}(a_{2i}^{\text{ga}})^2 
&
\sum_{i,-\frac{d}{\lambda} \sin \phi_{2i} \in B_2}^{}(a_{2i}^{\text{ga}})^2
&\ldots & 
\sum_{i,-\frac{d}{\lambda} \sin \phi_{2i} \in B_M}^{}(a_{2i}^{\text{ga}})^2
\\
\vdots \! & \! \vdots \! & \! \ddots \! & \! \vdots \!
\\
\sum_{i,-\frac{d}{\lambda} \sin \phi_{Ki} \in B_1}^{}(a_{Ki}^{\text{ga}})^2 
&
\sum_{i,-\frac{d}{\lambda} \sin \phi_{Ki} \in B_2}^{}(a_{Ki}^{\text{ga}})^2
&\ldots & 
\sum_{i,-\frac{d}{\lambda} \sin \phi_{Ki} \in B_M}^{}(a_{Ki}^{\text{ga}})^2
\end{bmatrix},
\label{U}
\end{split}
\end{IEEEeqnarray}
\end{figure*}
As (\ref{s2}) shows, the discrete-time Fourier transform at a particular $B_b$, is summation of the paths with DoDs in the same bin, i.e. $-\frac{d}{\lambda} \sin \phi_{kp} \in B_b$. Hence,
the estimated DoD based on the channel eigenvalue spectrum is not accurate for the case of finite $M$.
However, as we show in the next section, (\ref{s2}) can still be used to build up an``approximate eigenchannels'' matrix which can be used for beamforming and user scheduling. Note that by comparing  (\ref{s1}) and (\ref{s2}), we may conclude that by having a larger number of antennas the DoD of paths can be estimated perfectly, which demonstrates the effect of increasing the number of antennas at the BS.

In this paper, we evaluate the performance of collocated Massive MIMO for a realistic COST channel model. A possible, alternative system model is distributed Massive MIMO. Distributed Massive MIMO \cite{ourjournal2,ourjournal1,ourvtc18,ouricc2,our_lett_18,our_tgcn_accepted} with COST channel model has not been investigated before, but is out of the scope of this paper.

\section{Proposed User Scheduling and Beamforming}
In this paper, we aim to solve the problem of joint user scheduling and beamforming design assuming that only the second order statistics of the channel are available at the BS.
The proposed user selection scheme relies on 
a trade off between the number of occupied spectral bins for each user and the spectral overlap among the selected users. For this case, the performance analysis are found in the next subsection. Once the set of active users has been determined, the BS exploits the covariance matrix of the selected users for beamforming design and transmits data to the users.
\subsection{Correlation-based User Scheduling}\label{sub_coor_user}
By using the discrete-time Fourier transform of the antenna correlation given in (\ref{s2}), we generate the $K \times B$ matrix $\textbf{U}$ as (\ref{U}), where each $(k,b)$-th entry of the matrix $\textbf{U}$ denotes the discrete-time Fourier transform of the antenna correlation function of  the $k$th user at the $b$th bin, i.e. $\sum_{i,-\frac{d}{\lambda} \sin \phi_{ki} \in B_b}^{}(a_{ki}^{\text{ga}})^2 $.
The BS uses the functions $f_1(\textbf{u}_k)$ and $f_2(\textbf{u}_k)$ to perform user scheduling, where $\textbf{u}_k$ is the $k$th row of matrix ${\textbf{U}}$ and we define the functions $f_1(\textbf{u})$ and $f_2(\textbf{u})$ in the following.
As described in step $4.1$ in Algorithm \ref{al1}, the algorithm starts by calculating the summation over all area in terms of eigenvalue spectrum for all users, i.e. $f_1(\parallel\textbf{u}_k\parallel|) = \parallel\textbf{u}_k\parallel,~ \forall ~k$, and selects the user which has the largest value among the users. Then in the next step, the proposed algorithm finds a set of $\epsilon$-orthogonal users to the selected users.
Here, $\epsilon$-orthogonality among the user $k$ and the user $j$ means that $f_2(\textbf{u}_k,{\textbf{u}}_{j})=\frac{ |\textbf{u}_k{\textbf{u}}_{j}^{*}| }{ ||\textbf{u}_k||||{\textbf{u}}_{j}|| }<\epsilon$. Note that if the user $k$ and the user $j$ do not have spectral overlap, which means they do not have any shared bins, we have $\frac{  |\textbf{u}_k{\textbf{u}}_{j}^{*}| }{ ||\textbf{u}_k||||{\textbf{u}}_{j}|| }=0$. Hence, increasing the value of $\epsilon$ allows the users to have a bigger spectral overlap area. If the value of $\epsilon$ is too small, the area of spectral overlap between the selected users decreases and Algorithm \ref{al1} selects a small number of users. If the value of $\epsilon$ is too big, Algorithm \ref{al1} selects users with a large spectral overlap which can reduce the throughput  due to interference. It is well known that in GSCMs, MPCs from shared clusters cause high correlation which reduces the rank of the channel \cite{burrijas,alister_iswcs}. However, selecting users with no spectral overlap does not necessarily result in a higher throughput. So, to find the optimum value of $\epsilon$, we draw the sum rate versus $\epsilon$ and set the optimum value as $\epsilon$ in Algorithm \ref{al1}. Note that, $\mathcal{S}_{0}$ contains $K_s=|\mathcal{S}_{0}|$ indices of the selected users.
\begin{algorithm}[t!]
\caption{{Correlation-based user scheduling and beamforming (CUSBF):}}
\textbf{Step 1)} Initialization:~~${\Upsilon}_{0}=[1,\cdots,K]$, $\mathcal{S}_{0}=\emptyset$, $i=1$,\\
\textbf{Step 2)} Calculate the eigenvalue spectrum of $\textbf{R}_k$ by means of
the discrete-time Fourier transform of the antenna correlation
function,\\
\textbf{Step 3)} Generate matrix ${\textbf{U}}$ given by (\ref{U}),\\
\textbf{Step 4)} Greedy Algorithm:
\begin{itemize}
\item{\textbf{\!\!\!4.1}}
$\pi(i)=\argmax_{k \in \Upsilon_{0}}f_1(\parallel\textbf{u}_k\parallel)\\=\argmax_{k \in \Upsilon_{0}} \parallel\textbf{u}_k\parallel$, $\mathcal{S}_{0}\gets \mathcal{S}_{0}\cup \{k\}$, ${\textbf{u}}_{(i)}=\textbf{u}_{(\pi(i))}$,\\
\item{\textbf{\!\!\!4.2}} If $|\Upsilon_{0}|<K_s$, ${\Upsilon}_{i}=\{k \in {\Upsilon}_{i-1}, k \ne \pi(i) \mid f_2(\textbf{u}_k,{\textbf{u}}_{(i)})=\frac{  |\textbf{u}_k{\textbf{u}}_{(i)}^{*}|  }{ \parallel\textbf{u}_k\parallel\parallel{\textbf{u}}_{(i)}\parallel  }<\epsilon \}$,\\
\item{\textbf{\!\!\!4.3}}
If $|\Upsilon_0|<K_s$ and $\Upsilon_{i}\neq\emptyset$, then $i \leftarrow i+1$, and go to step 4.1, else, go to step 5,
\end{itemize}
\textbf{Step 5}) Generate matrix ${\textbf{G}}$ given by (\ref{htilde}). BS does not require the instantaneous channels of the users and uses matrix $\textbf{G}$ for beamforming design.
\label{al1}
\end{algorithm}
\subsection{Correlation-based Beamforming}
Once the set of users is fixed, the BS can design the precoding matrix based on the knowledge of $\textbf{R}_k,~\forall k$. If $\textbf{R}_k,~\forall k$, is available at the BS, it is possible to find an approximated version for the channel matrix $\textbf{G}$. So, at step 5 of Algorithm \ref{al1}, we propose to build up the approximate eigenchannel matrix for the channels of users based on the channel covariance matrix given by eq. (\ref{s2}) as follows:

\begin{IEEEeqnarray}{rCl}
{g}_{km}=
\sum_{b=1}^{M}\!\left(\!\sum_{i, -\frac{d}{\lambda}\sin \phi_{ki}\in {\text{B}_b}}^{N_l}\left(a_{ki}^{ga}\right)^2\!\right)^{\frac{1}{2}}\!
\!e^{j2\pi\left(m-1\right)\left(\frac{b}{M}-\frac{1}{2}\!\right)}\!,~~
\label{hfft}
\end{IEEEeqnarray}
\begin{figure*}[!t]
\begin{IEEEeqnarray}{rCl}
\text{Finite}~M;
\begin{split}
{\textbf{G}}&\!=\!&\!\begin{bmatrix}\!\sum_{b=1}^{M}(\sum_{i, -\frac{d}{\lambda}\sin \phi_{1i}\in {\text{B}_b}}^{N_l}(a_{1i}^{\text{ga}})^2)^{\frac{1}{2}}\!
\!&\! \ldots \!&\!
\sum_{b=1}^{M}(\sum_{i, -\frac{d}{\lambda}\sin \phi_{1i}\in {\text{B}_b}}^{
N_l}(a_{1i}^{\text{ga}})^2)^{\frac{1}{2}}e^{j2\pi(M-1)(\frac{b}{M}-\frac{1}{2})}\!
\\
\sum_{b=1}^{M}(\sum_{i, -\frac{d}{\lambda}\sin \phi_{2i}\in {\text{B}_b}}^{
N_l}(a_{2i}^{\text{ga}})^2)^{\frac{1}{2}}\!
\!&\!\ldots\!&\!
\sum_{b=1}^{M}(\sum_{i, -\frac{d}{\lambda}\sin \phi_{2i}\in {\text{B}_b}}^{
N_l}(a_{2i}^{\text{ga}})^2)^{\frac{1}{2}}e^{j2\pi(M-1)(\frac{b}{M}-\frac{1}{2})}\!
\\\vdots & \ddots & \vdots \\
\sum_{b=1}^{M}(\sum_{i, -\frac{d}{\lambda}\sin \phi_{Ki}\in {\text{B}_b}}^{
N_l}(a_{Ki}^{\text{ga}})^2)^{\frac{1}{2}}\!
\!&\!\ldots \!&\!
\sum_{b=1}^{M}(\sum_{i, -\frac{d}{\lambda}\sin \phi_{Ki}\in {\text{B}_b}}^{
N_l}(a_{Ki}^{\text{ga}})^2)^{\frac{1}{2}}e^{j2\pi(M-1)(\frac{b}{M}-\frac{1}{2})}
\end{bmatrix}.
\label{htilde}
\end{split}
\end{IEEEeqnarray}
\hrulefill
\end{figure*}
where the approximate eigenchannel ${g}_{km}$ is a superposition of $B$ approximated paths, where $B=M$ (denotes the total number of angular bins) and the $b$th approximated path is centred at $\frac{b}{M}-\dfrac{1}{2}$. We propose that the BS uses equation (\ref{hfft}) to build up the approximate eigenchannel matrix ${\textbf{{G}}}$ defined in (\ref{htilde}) at the top of the next page.
The approximate eigenchannel matrix ${\textbf{{G}}}$ can be used for user scheduling and precoding design. Note that only for the case of $M\to \infty$, the DoD of each single MPC is resolvable and are available at the BS. The investigation of the proposed scheme with the relay-assisted \cite{mehdi1,mehdi2,mehdi3} Massive MIMO will be considered in our future work. 
\begin{table}[!t] \centering \caption{Computational Complexity of Different Schemes}
\label{tab:ComparisonTime} 
\begin{IEEEeqnarraybox}[\IEEEeqnarraystrutmode\IEEEeqnarraystrutsizeadd{2pt}{1pt}]{x;s;x;s;x;t;x;t;x;t;x;t;x;t;x} 
\IEEEeqnarrayrulerow[.11em]\\ 
 &\textbf{Schemes}&&\!\!\!\!\!\!\!\!\!\!\!\textbf{Channel estimation}&&\textbf{user Scheduling}&& \textbf{Beamforming~~~~~~}\\ 
\IEEEeqnarrayrulerow[.2em]\vspace{.11cm} \\ 
 &\textbf{GWC \cite{SUSGoldsmithGlobcom}}&&$\mathcal{O}(K^3M^3)$&&~~~~~~$\!\!\!\!\!\!\!\!\!\!\!\!\!\!\!\!\!\!\!\!\!\!\!\!\!\mathcal{O}(K)$&&$\mathcal{O}(M^3)~~~~~~~~~$\\ 
\IEEEeqnarrayrulerow[.17em]\vspace{.11cm} \\ 
 &\textbf{JSDM \cite{CiareJointspatial}}&&$\mathcal{O}(K_s^3M^3)$&&~~~~~~$\!\!\!\!\!\!\!\!\!\!\!\!\!\!\!\!\!\!\!\!\!\!\!\!\!\!\mathcal{O}(K)$&&
 $\!\!\!\!\!\!\!\!\!\!\!\!\!\!K_s\mathcal{O}\!\left(\!M^3 \!\!+\!\! M  \log^2M  \log b\right)~~~~~~$\\
\IEEEeqnarrayrulerow[.2em] \vspace{.11cm} \\
 &\textbf{\small{Algorithm 1}}&&$~~~~-~~~~$&&~~~~~~$\!\!\!\!\!\!\!\!\!\!\!\!\!\!\!\!\!\!\!\!\!\!\!\!\!\mathcal{O}(K)$&&$\mathcal{O}(M^3)~~~~~~~~~$\\ 
\IEEEeqnarrayrulerow[.18em]\\\
&&&&&\IEEEeqnarraymulticol{9}{c}{~~~~~~~~~~~~~~~~~~~~~~~~~~~~~~~~~~~~~~~~~~~~~~~~}&\\ 
\end{IEEEeqnarraybox}
\label{complex}
\end{table}
\section{Complexity Analysis}
Without loss of
generality the complexity, computation of the minimum mean square error (MMSE) estimator is given by $\mathcal{O}(\tau^3  M^3)$, where $\tau=K$ is sufficient to remove the effect of pilot contamination \cite{debbah_cor_es}. Hence, the complexity of the MMSE estimator scales as $\mathcal{O}(K^3  M^3)$, which indicates the complexity of inverting of matrix size $KM \times KM$ to estimate the channel in equation (\ref{h3}), which is required in the GWC scheme in \cite{SUSGoldsmithGlobcom}. The proposed Algorithm 1 and the JSDM scheme in \cite{CiareJointspatial} do not exploit the knowledge of channel for user scheduling and beamforming design. For a given $M\times M$ matrix,
the required operations to determine the eigenvectors
is given by $\mathcal{O}\big[M^3 + (M  \log^2M)  \log b\big]$, where $b$ is the relative error bound \cite{bookeigen}. Moreover, the complexity to search the user for the scheme in \cite{SUSGoldsmithGlobcom} is linear with the number of users  \cite{Yoothesis}. Note that the complexity of user scheduling in the proposed Algorithm 1 and the scheme in \cite{CiareJointspatial,mythesis} is linear in terms of the number of users. The number of arithmetic operations required for Algorithm \ref{al1} is shown in Table \ref{complex}. The authors in \cite{iraconmeeting_correlation} define the spatial WSS quality which is given by
\begin{IEEEeqnarray}{rCl}
	Q_{\text{WSS}}=\frac{\tau_{\text{LT}}}{\tau_c},
\end{IEEEeqnarray}
where $\tau_{\text{LT}}$ refers to the long-term time, where the statistics of the channel may be considered constant within this interval whereas $\tau_c$ is the channel coherence time. The measurement results for the outdoor scenario at a center frequency of 2 GHz shows that $Q_{\text{WSS}}=120$. As a result, every $120\times \tau_c$, the correlation based schemes (the proposed Algorithm 1 and the scheme in \cite{CiareJointspatial}) need to be run, while the scheme in \cite{SUSGoldsmithGlobcom} need to be run at the beginning of each coherence time.

\section{Numerical Results and Discussion}
\label{num} 
A square cell with a side length of $2\times R$ has been considered; we call $R$ the cell size and also assume users are uniformly distributed in the cell. As in \cite{MarzettaMRC13}, we assume that there is no user closer than $R_{th}=0.1\times R$ to the BS. We simulate a micro-cell environment for the NLoS case and set the operating frequency $f_C=2$ GHz. The external parameters and stochastic parameters are extracted from chapter 3 of \cite{Costaction}. The BS and user heights are assumed to be $h_{BS}=5$ m and $h_{MS}=1.5$ m, respectively. 
The noise power is given by
$
P_n = \text{BW}~k_B~T_0~W,
$
where $\text{BW}=20$ MHz denotes the bandwidth, $k_B = 1.381 \times 10^{-23}$ represents the Boltzmann constant, $T_0 = 290$ (Kelvin) denotes the noise temperature, and $W=9$ dB is the noise figure.
\begin{figure*}%
\centering
\begin{subfigure}{.683\columnwidth}
\includegraphics[width=\columnwidth]{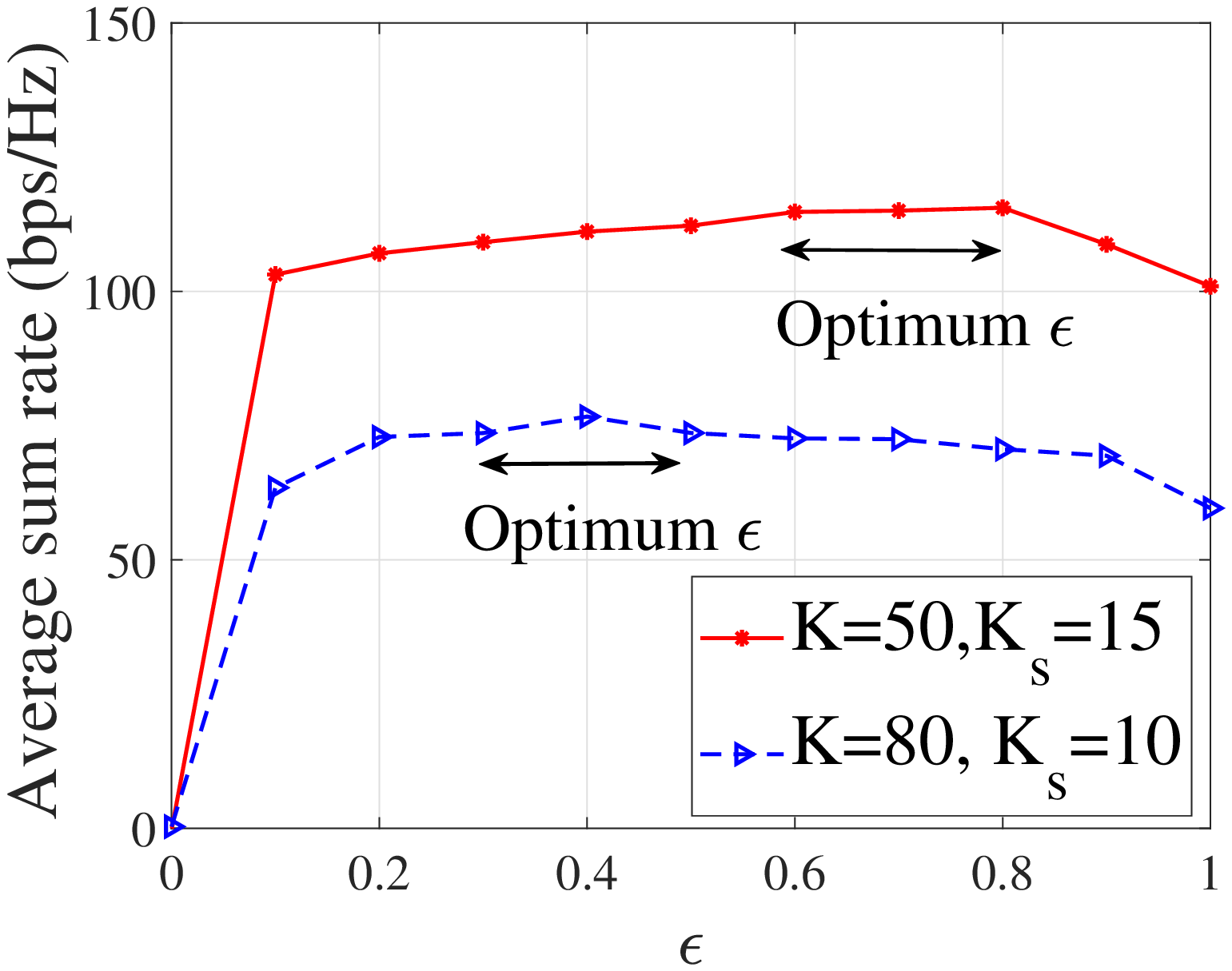}%
\caption{Average sum rate versus $\epsilon$.}%
\label{subfiga}%
\end{subfigure}\hfill%
\begin{subfigure}{.67\columnwidth}
\includegraphics[width=\columnwidth]{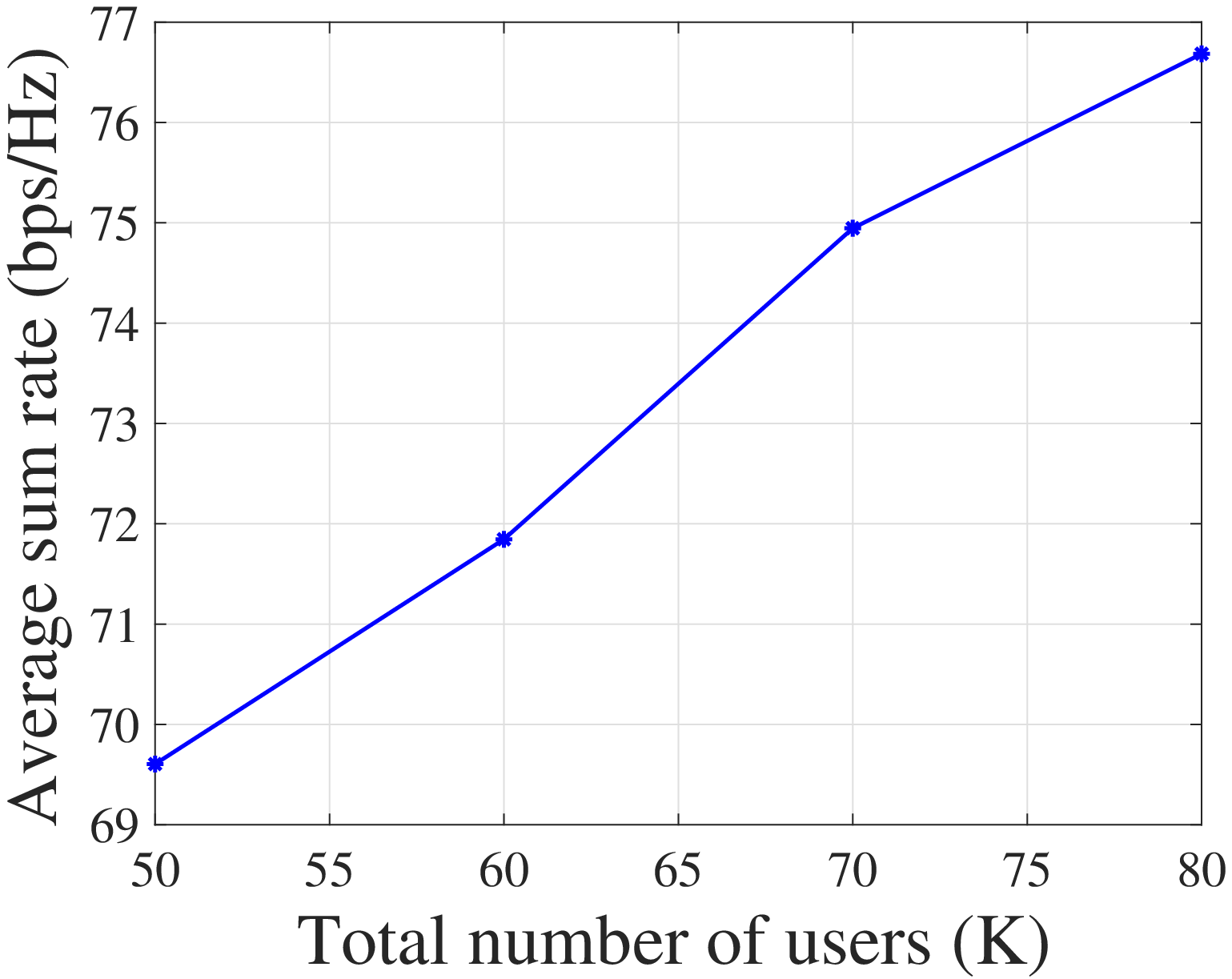}%
\caption{Average sum rate versus total number of\\ users with $K_s=10$.}%
\label{subfigb}%
\end{subfigure}\hfill%
\begin{subfigure}{.67\columnwidth}
\includegraphics[width=\columnwidth]{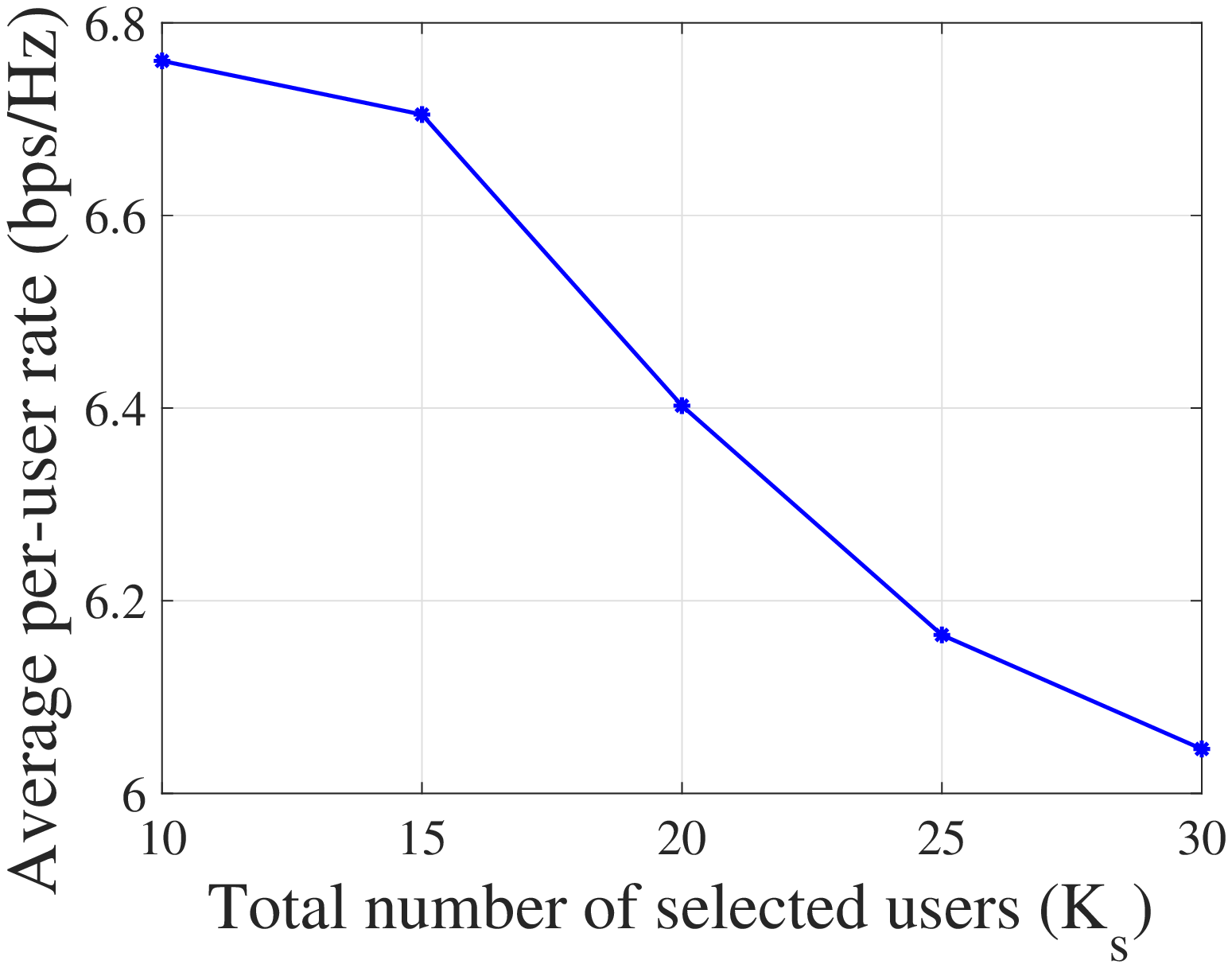}%
\caption{Average per-user rate versus total number of selected antennas with $K=50$.}%
\label{subfigc}%
\end{subfigure}%
\caption{The performance of Algorithm 1 with $p_k=10$ dBm and $R=500$ meters.}
\label{figabc}
\end{figure*}
\begin{figure}[t!]
\center
\includegraphics[width=90mm]{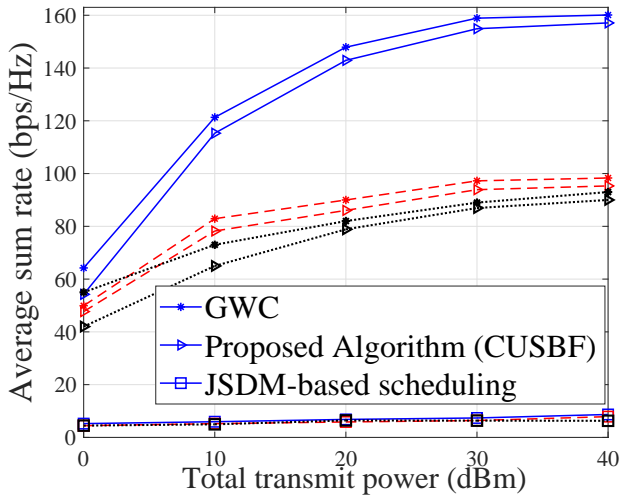}
\caption{The average sum rate vs. transmit power. Solid (blue), dashed (red) and dotted (black) lines refer to $\{M=300, K=70,K_s=20\}$, $\{M=300, K=50,K_s=10\}$ and $\{M=200, K=50,K_s=10\}$, respectively.}
\label{gold_prop_cair_m300_m200_k70_k50_ks20_ks10}
\end{figure}
For this network setup, the average sum rate is evaluated for the three scenarios. First, we evaluate the average throughput of the proposed Correlation-based user scheduling and beamforming (CUSBF) scheme, given by Algorithm 1.
In Fig. \ref{subfiga}, the sum rate of users under the proposed scheme 
is plotted as a function of $\epsilon$ in Algorithm 1. If $\epsilon$ is too large, the spectral overlap (number of shared bins) is big, while if is too small, the multiuser diversity gain decreases and users with shared bins cannot be selected. As a result, there should be a trade off between total number of shared bins and summation over all area in terms of eigenvalue spectrum, which is explained in Subsection \ref{sub_coor_user}.
The optimal value of $\epsilon$ is shown in Fig. \ref{subfiga}. Next, we plot the average sum rate versus the total number of users in the system in Fig. \ref{subfigb}. As the figure shows, by increasing the total number of users, the average sum rate increases, as a result of multi-user diversity gain. Fig. \ref{subfigc} demonstrates the average per-user rate versus the total number of users in the system. Note that the analysis in \cite{scaleup} demonstrate that in the limit of Massive MIMO ($M ,K_s\to \infty $ and $\alpha=\frac{M}{K_s}$), by increasing $K_s$ the average per-user rate decreases.

Finally, we evaluate the average throughput of the proposed CUSBF scheme, given by Algorithm 1, and GWC \cite{SUSGoldsmithGlobcom,ITC09_Userselection_GWC} with an MMSE estimate of the channel. For the case of GWC, similar to \cite{ITC09_Userselection_GWC}, we set the optimal channel direction
constraint to achieve the best performance for GWC. Moreover, the comparison with the scheme proposed in \cite{CiareJointspatial} is provided. In \cite{CiareJointspatial}, Adhikari et al. propose to select users which occupy a larger number of bins and find users having a smaller spectral overlap with the selected users. This scheme is referred to JSDM-based scheduling. 
The analysis in \cite{scaleup} demonstrates that in the limit of Massive MIMO ($M ,K_s \to \infty $ and $\kappa=\frac{M}{K_s}$),
when $\kappa\ge 5$, linear precoding is “virtually optimal”, 
and can be used instead of dirty paper coding (DPC).
In this paper, we follow the network setup introduced in \cite{CiareJointspatial} and \cite{MarzettaMRC13}, and we choose $\kappa=15$ and $\kappa=30$. This is given by two cases with $\kappa=\frac{300}{20}=15$ and $\kappa=\frac{300}{10}=30$, respectively. Moreover, note that assuming 20 users and a BS having 300 antennas at a frequency of 2 GHz is common assumption \cite{Sadeghi_multigroup_18}.

Fig. \ref{gold_prop_cair_m300_m200_k70_k50_ks20_ks10} depicts the average sum rate versus the total transmit power for three cases of $\{M=300, K=70,K_s=20\}$, $\{M=300, K=50,K_s=10\}$ and $\{M=200, K=50,K_s=10\}$, while adopting the currently proposed scheme with zero-forcing beamforming (ZFBF). As expected, since GWC exploits the estimated instantaneous CSI, it has the best throughput. In addition, the figure demonstrates that in the medium and high SINR regime the difference between the proposed CUSBF scheme and the GWC scheme is smaller.
As the figures show, the performance of the proposed Algorithm 1 is slightly poorer than the case in which the BS has the knowledge of the estimated instantaneous channel to perform user scheduling and beamforming as in \cite{SUSGoldsmithGlobcom}, i.e., GWC. Interestingly, for a larger number of antennas at the BS, the superiority of the proposed scheme is more obvious in terms of achieving performance close to that of the GWC scheme. Moreover, the performance of the proposed algorithm is several times higher than for the scheme in \cite{CiareJointspatial}, i.e., JSDM-based scheduling. In addition, the figure demonstrate that the performance of the scheme in \cite{CiareJointspatial} is quite poor for the case of the COST 2100 channel model. This is because of the large number of clusters in the area, which means that the performance of eigen-beamforming is not as good as ZFBF. Note that the JSDM in \cite{CiareJointspatial} is designed to work well with the angularly-sparse multipath channels typically observed in the mm-waves. 
\vspace{-.23cm}
\section{Conclusions}
We proposed to use the angular bins of the eigenvalue spectrum of the channel covariance matrix to build up an approximate eigenchannel for the users. Using the discrete-time Fourier transform of the antenna correlation function, a novel user scheduling scheme and linear precoding design has been proposed and tested with the COST 2100 channel model.
The results show that while the average throughput slightly decreases due to absence of instantaneous channel, the computational complexity of the system reduces significantly. As a result, the proposed scheme can be considered as a superior practical approach for massive MIMO systems.
\vspace{-.2cm}
\bibliographystyle{IEEEtran}
\bibliography{TVT_final_rose} 
\end{document}